\documentclass[twoside,english,10pt,journal,a4paper,twocolumn]{IEEEtran}
\usepackage[T1]{fontenc}
\usepackage[latin9]{inputenc}
\usepackage{textcomp}
\usepackage{amsmath}
\usepackage{amssymb}
\usepackage{graphicx}

\makeatletter

\providecommand{\tabularnewline}{\\}

\usepackage{changebar}

\@ifundefined{showcaptionsetup}{}{%
 \PassOptionsToPackage{caption=false}{subfig}}
\usepackage{subfig}
\makeatother

\begin{document}

\title{Maximizing Energy-Efficiency in Multi-Relay OFDMA Cellular Networks}

\author{Kent Tsz Kan Cheung, Shaoshi Yang and Lajos Hanzo\thanks{This research has been funded by the Industrial Companies who are Members of the Mobile VCE, with additional financial support from the UK Government's Engineering \& Physical Sciences Research Council (EPSRC). The financial support of the China Scholarship Council (CSC), of the Research Councils UK (RCUK) under the India-UK Advanced Technology Center (IU-ATC), and of the EU under the auspices of the Concerto project is also gratefully acknowledged. Finally, the fiscal support of the European Research Council under its Advanced Fellow Grant is thankfully acknowledged.}\\
{\small{} School of ECS, University of Southampton, SO17 1BJ, United
Kingdom.}\\
{\small{} Email: \{ktkc106,sy7g09,lh\}@ecs.soton.ac.uk, http://www.cspc.ecs.soton.ac.uk}\vspace{-0.5cm}
}

\markboth{in 2013 IEEE 56th Global Communications Conference (GLOBECOM 2013),~Atlanta,~USA,~December~2013} %
{Shell \MakeLowercase{\textit{et al.}}: Bare Demo of IEEEtran.cls
for Journals}

\maketitle
\begin{abstract}
This contribution presents a method of obtaining the optimal power
and subcarrier allocations that maximize the energy-efficiency~(EE)
of a multi-user, multi-relay, orthogonal frequency division multiple
access~(OFDMA) cellular network. Initially, the objective function~(OF)
is formulated as the ratio of the spectral-efficiency~(SE) over the
power consumption of the network. This OF is shown to be quasi-concave,
thus Dinkelbach's method can be employed for solving it as a series
of parameterized concave problems. We characterize the performance
of the aforementioned method by comparing the optimal solutions obtained
to those found using an exhaustive search. Additionally, we explore
the relationship between the achievable SE and EE in the cellular
network upon increasing the number of active users. In general, increasing
the number of users supported by the system benefits both the SE and
EE, and higher SE values may be obtained at the cost of EE, when an
increased power may be allocated.
\end{abstract}

\section{Introduction\label{sec:Introduction}}

In recent years, increasing the energy-efficiency~(EE) of cellular
networks has become an important design metric in the telecommunications
community, especially in the light of the conflicting criteria of
achieving both an increased data rate as well as reducing the 'carbon
footprint'. Consequently, joint academic and industrial effort has
been dedicated to developing novel energy-saving techniques for next-generation
networks~\cite{Han2011}. This contribution considers the downlink~(DL)
EE maximization~(EEM) problem in a multi-user, multi-relay, orthogonal
frequency division multiple access~(OFDMA) network such as that specified
both in the third generation partnership project's~(3GPP) long term
evolution-advanced~(LTE-A) and in the IEEE 802.16 worldwide interoperability
for microwave access~(WiMAX) standards.

Numerous contributions have already dealt with the problem of allocating
power and/or subcarriers in an OFDMA network with the goal of either
spectral-efficiency~(SE) maximization~\cite{Ng2010}, or power minimization
subject to specific quality-of-service~(QoS) constraints~\cite{Wong1999,Joung2012}.
However, these works have not solved the EEM problem, which has only
recently become the center of attention in the community. In fact,
the EEM problem can be viewed as an example of multi-objective optimization,
since typically the goal is to maximize the SE achieved, while concurrently
minimizing the power consumption. Hence the authors of~\cite{Devarajan2012}
derived an aggregate OF, which consists of a weighted sum of the SE
achieved and the total power dissipated. However, selecting appropriate
weights for the two OFs is not trivial, and different combinations
of weights can lead to different results. Another example is given
in~\cite{Yu2011}, where the EEM problem is considered in a multi-relay
network. Nonetheless, both~\cite{Devarajan2012,Yu2011} only optimize
the user selection and power allocation without considering the subcarrier
allocation in the network. Another formulation was advocated in~\cite{Miao2012},
which considers the power and subcarrier allocation problem in an
OFDMA cellular network, but without a maximum total power constraint
and without relaying. The authors of~\cite{Ng2012} formulated the
EEM problem in a OFDMA cellular network under a maximum total power
constraint, but relaying was not considered.

In contrast to the aforementioned related work, this contribution
focuses on jointly optimizing both the power and subcarrier allocation
for a multi-relay, multi-user OFDMA cellular network, while considering
a specific maximum total power constraint. As a benefit of its low-complexity
implementation, we employ the amplify-and-forward~(AF)~\cite{Laneman2004}
protocol at the relays. The contributions of this paper are summarized
as follows.

The EEM problem, in the context of a multi-relay, multi-user OFDMA
cellular network, in which both direct and relayed transmissions are
employed, is formulated as a fractional programming problem, which
jointly considers both the power and the subcarrier allocation under
a maximum power constraint. This problem is relaxed%
\footnote{In the context of optimization, relaxing a problem is equivalent to
forming the same problem but with looser constraints.%
} and can be shown to be quasi-concave, therefore Dinkelbach's method~\cite{Dinkelbach1967}
may be employed for iteratively obtaining the optimal solution. It
is demonstrated that the EEM algorithm reaches the optimal solution
within a low number of iterations and succeeds in reaching the solution
obtained via an exhaustive search. Thus the original problem is solved
at a low complexity.

Furthermore, comparisons are made between the EEM and the SE maximization~(SEM)
based solutions. As an example, it is shown that when the maximum
affordable power is lower than a given threshold, the two problems
have the same solutions. However, as the maximum affordable power
is increased, the SEM algorithm attempts to achieve a higher SE at
the cost of a lower EE. By contrast, given the total power, the EEM
algorithm reaches the upper limit of the maximum achievable SE for
the sake of maintaining the maximum EE. Additionally, we will demonstrate
that both the SE and EE benefit from an increased multi-user diversity
in the network.

\section{System Model\label{sec:System-Model}}

\begin{figure}
\centering{}\includegraphics[scale=0.5]{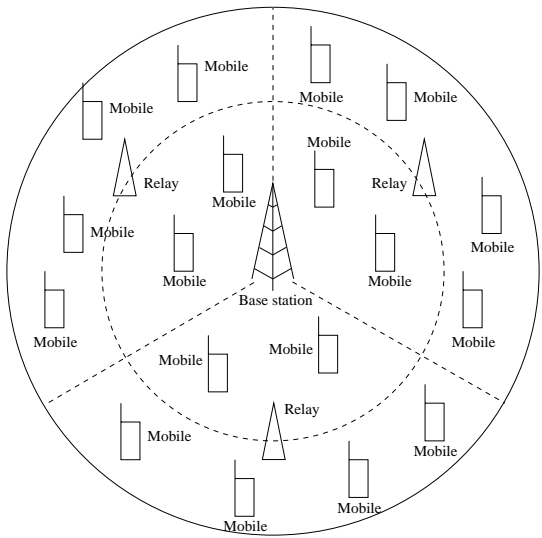}\caption{An example of a cellular network with $M=3$ RNs and $K=18$ UEs.}
\label{fig:cellular}
\end{figure}

We consider an OFDMA DL cellular system consisting of a single base
station~(BS), $M$ fixed relay nodes~(RNs) and $K$ uniformly-distributed
user-equipment~(UEs), as shown in Fig.~\ref{fig:cellular}. This
network has access to $N$ subcarriers, and the cell is partitioned
into $M$ sectors such that the UEs in each sector are served by the
single RN. In order to reduce the detrimental effect of path-loss
on the achievable SE and EE, each UE may only be served by the specific
RN that it is closest to, and thus RN selection is implicitly accomplished.
The BS may perform DL transmissions either via a direct BS-to-UE link,
or by relying on the RN for creating an AF BS-to-RN-to-UE link. Additionally,
we assume that the total available instantaneous transmission power
of the network is $P_{max}$. Below, we adopt the following notation.
The variables related to the two possible communication protocols
are denoted by the superscripts $D$ and $A$, respectively. When
defining links, the subscript $0$ is used for indicating the BS,
while $\mathcal{M}(k)\in\{1,\cdots,M\}$ indicates the RN selected
for assisting the DL-transmissions to user $k$.

The attained SE when using the direct and relayed transmissions~\cite{Laneman2004}
may be expressed as
\begin{equation}
R_{k}^{D,n}(\mathcal{P})=\mbox{log}_{2}\left(1+\frac{P_{0,k}^{D,n}G_{0,k}^{n}}{\Delta\gamma N_{0}W}\right)\mbox{ }\mbox{[bits/s/Hz]}\label{eq:R_D}
\end{equation}
and~(\ref{eq:R_A}), respectively.\begin{figure*}[tbp]
\begin{equation}
R_{k}^{A,n}(\mathcal{P})\approx\frac{1}{2}\mbox{log}_{2}\left(1+\frac{P_{0,\mathcal{M}(k)}^{A,n}G_{0,\mathcal{M}(k)}^{n}P_{\mathcal{M}(k),k}^{A,n}G_{\mathcal{M}(k),k}^{n}}{\Delta\gamma N_{0}W\left(P_{0,\mathcal{M}(k)}^{A,n}G_{0,\mathcal{M}(k)}^{n}+P_{\mathcal{M}(k),k}^{A,n}G_{\mathcal{M}(k),k}^{n}\right)}\right)\mbox{ }\mbox{[bits/s/Hz]},\label{eq:R_A}
\end{equation}
\hrulefill
\end{figure*} To elaborate, $P_{a,b}^{X,n}$ is the power allocated to transmitter
$a\in\{0,\cdots,M\}$ using protocol $X\in\{D,A\}$ for transmission
to receiver $b$ on subcarrier $n$. Furthermore, $G_{a,b}^{n}$ represents
the channel's attenuation between transmitter $a$ and receiver $b$
on subcarrier $n$, which is assumed to be known at the BS for all
links, $N_{0}$ is the additive white Gaussian noise~(AWGN) variance
and $W$ is the bandwidth of a single subcarrier. The signal-to-noise~(SNR)
gap, $\Delta\gamma$, is measured at the system's bit error ratio~(BER)
target, and is the difference between the SNR required at the discrete-input
continuous-output memoryless channel~(DCMC) capacity and the actual
SNR required the modulation and coding schemes of the practical physical
layer transceivers employed. In Section~\ref{sec:Results-and-Discussions},
we make the simplifying assumption that idealized transceivers operating
exactly at the DCMC capacity are employed, thus $\Delta\gamma=0\mbox{ dB}$.
Additionally, the power allocation policy of the system is denoted
by $\mathcal{P}$, which determines the values of $P_{a,b}^{X,n}$.
The factor of $\frac{1}{2}$ in~(\ref{eq:R_A}) accounts for the
fact that only half the transmission time is available for transmitting
new data, while the other half must be used for the AF relaying. Additionally,
high receiver's SNR values is assumed in~(\ref{eq:R_A}), which is
valid for $P_{0,\mathcal{M}(k)}^{A,n}G_{0,\mathcal{M}(k)}^{n}+P_{\mathcal{M}(k),k}^{A,n}G_{\mathcal{M}(k),k}^{n}\gg\Delta\gamma N_{0}W$.

The subcarrier indicator variable $s_{k}^{X,n}\in\{0,1\}$ is now
introduced, which denotes the allocation of subcarrier $n$ for transmission
to user $k$ using protocol $X$ when $s_{k}^{X,n}=1$, and $s_{k}^{X,n}=0$
otherwise. The average SE of the system is then calculated as
\begin{equation}
R_{T}(\mathcal{P},\mathcal{S})=\frac{1}{N}\sum_{k=1}^{K}\sum_{n=1}^{N}s_{k}^{D,n}R_{k}^{D,n}+\frac{s_{k}^{A,n}}{2}R_{k}^{A,n}\mbox{\mbox{ }[bits/s/Hz]},
\end{equation}
where $\mathcal{S}$ denotes the subcarrier allocation policy of the
system, which determines the values of the subcarrier indicator variable
$s_{k}^{X,n}$.

In order to compute the energy used in these transmissions, a model
similar to~\cite{Arnold2010} is adopted where the total power consumption
of the system is assumed be governed by a constant term and a term
that varies with the transmission powers, which may be written as~(\ref{eq:P_T}).\begin{figure*}[tbp]
\begin{equation}
P_{T}(\mathcal{P},\mathcal{S})=\left(P_{C}^{(B)}+M\cdot P_{C}^{(R)}\right)+\sum_{k=1}^{K}\sum_{n=1}^{N}s_{k}^{D,n}\xi^{(B)}P_{0,k}^{D,n}+\frac{1}{2}s_{k}^{A,n}\cdot\left(\xi^{(B)}P_{0,\mathcal{M}(k)}^{A,n}+\xi^{(R)}P_{\mathcal{M}(k),k}^{A,n}\right)\mbox{ }\mbox{[Watts]}\label{eq:P_T}
\end{equation}
\hrulefill
\end{figure*} Here, $P_{C}^{(B)}$ and $P_{C}^{(R)}$ represent the fixed power
consumption of each BS and each RN, respectively, while $\xi^{(B)}>1$
and $\xi^{(R)}>1$ denote the reciprocal of the drain efficiencies
of the power amplifiers employed at the BS and the RNs, respectively.
For example, an amplifier having a $25\%$ drain efficiency would
have $\xi=\frac{1}{0.25}=4$.

Finally, the average EE metric of the system is expressed as
\begin{equation}
\eta_{E}(\mathcal{P},\mathcal{S})=\frac{R_{T}(\mathcal{P},\mathcal{S})}{P_{T}(\mathcal{P},\mathcal{S})}\mbox{ [bits/Joule/Hz]}.\label{eq:eeff}
\end{equation}

\section{Problem Formulation\label{sec:Problem-Formulation}}

The aim of this work is to maximize the EE metric of~(\ref{eq:eeff})
subject to a maximum total instantaneous transmit power constraint.
In its current form,~(\ref{eq:eeff}) is dependent on $3KN$ continuous
power variables $P_{0,k}^{D,n}$, $P_{0,\mathcal{M}(k)}^{A,n}$ and
$P_{\mathcal{M}(k),k}^{A,n}$, $\forall k,n$, and $2KN$ binary subcarrier
indicator variables $s_{k}^{D,n}$ and $s_{k}^{A,n}$, $\forall k,n$.
Thus, it may be regarded as a MINLP problem, and can be solved using
the branch-and-bound method~\cite{Bertsekas1999}. However, the computational
effort required for branch-and-bound techniques typically increases
exponentially with the problem size. Therefore, a simpler solution
is derived by relaxing the binary constraint imposed on the subcarrier
indicator variables, $s_{k}^{D,n}$ and $s_{k}^{A,n}$, so that they
may assume continuous values from the interval $[0,1]$, as demonstrated
in~\cite{Wong1999,Wei2006}. Furthermore, the auxiliary variables
$\widetilde{P}_{0,k}^{D,n}=P_{0,k}^{D,n}s_{k}^{D,n}$, $\widetilde{P}_{0,\mathcal{M}(k)}^{A,n}=P_{0,\mathcal{M}(k)}^{A,n}s_{k}^{A,n}$
and $\widetilde{P}_{0,\mathcal{M}(k)}^{A,n}=P_{0,\mathcal{M}(k)}^{A,n}s_{k}^{A,n}$
are introduced.

The relaxation of the binary constraints imposed on the variables
$s_{k}^{D,n}$ and $s_{k}^{A,n}$ allows them to assume continuous
values, which leads to a time-sharing subcarrier allocation between
the UEs. Naturally, the original problem is not actually solved. However,
it has been shown that solving the dual of the relaxed problem provides
solutions that are arbitrarily close to the original, non-relaxed
problem, provided that the number of available subcarriers tends to
infinity~\cite{Wei2006}. It has empirically been shown that in some
cases only $8$ subcarriers are required for obtaining close-to-optimal
results~\cite{Seong2006}. It shall be demonstrated in Section~\ref{sec:Results-and-Discussions}
that even for as few as two subcarriers, the solution algorithm employed
in this work approaches the optimal EE achieved by an exhaustive search.

The optimization problem is formulated as shown as follows.

Relaxed Problem (P):
\begin{eqnarray}
\underset{\mathcal{P},\mathcal{S}}{\mbox{maximize}} &  & \frac{\widetilde{R}_{T}}{\widetilde{P}_{T}}\label{eq:eeff_orig}\\
\mbox{}\nonumber \\
\mbox{ subject to} &  & \sum_{k=1}^{K}\sum_{n=1}^{N}\widetilde{P}_{0,k}^{D,n}+\widetilde{P}_{0,\mathcal{M}(k)}^{A,n}+\widetilde{P}_{\mathcal{M}(k),k}^{A,n}\leq P_{max},\nonumber \\
\label{eq:C1}\\
 &  & s_{k}^{D,n}+s_{k}^{A,n}\leq1\mbox{, }\forall k,n,\label{eq:C2}\\
 &  & \sum_{k=1}^{K}s_{k}^{D,n}+s_{k}^{A,n}\leq1\mbox{, }\forall n,\label{eq:C3}\\
 &  & \widetilde{P}_{0,k}^{D,n}\mbox{, }\widetilde{P}_{0,\mathcal{M}(k)}^{A,n}\mbox{, }\widetilde{P}_{\mathcal{M}(k),k}^{A,n}\in\mathbb{R}_{+}\mbox{, }\forall k,n,\label{eq:C4}\\
 &  & 0\leq s_{k}^{D,n}\mbox{, }s_{k}^{A,n}\leq1\mbox{, }\forall k,n,\label{eq:C5}
\end{eqnarray}
where the objective function is the ratio between~(\ref{eq:RR_T})
and~(\ref{eq:PP_T}).

\begin{figure*}[tbp]
\begin{eqnarray}
\widetilde{R}_{T} & = & \sum_{k=1}^{K}\sum_{n=1}^{N}s_{k}^{D,n}\mbox{log}_{2}\left(1+\frac{\widetilde{P}_{0,k}^{D,n}G_{0,k}^{n}}{s_{k}^{D,n}\Delta\gamma N_{0}W}\right)\nonumber \\
 &  & +\frac{s_{k}^{A,n}}{2}\mbox{log}_{2}\left(1+\frac{\widetilde{P}_{0,\mathcal{M}(k)}^{A,n}G_{0,\mathcal{M}(k)}^{n}\widetilde{P}_{\mathcal{M}(k),k}^{A,n}G_{\mathcal{M}(k),k}^{n}}{s_{k}^{A,n}\Delta\gamma N_{0}W\left(\widetilde{P}_{0,\mathcal{M}(k)}^{A,n}G_{0,\mathcal{M}(k)}^{n}+\widetilde{P}_{\mathcal{M}(k),k}^{A,n}G_{\mathcal{M}(k),k}^{n}\right)}\right)\label{eq:RR_T}
\end{eqnarray}
\begin{equation}
\widetilde{P}_{T}=\left(P_{C}^{(B)}+M\cdot P_{C}^{(R)}\right)+\sum_{k=1}^{K}\sum_{n=1}^{N}\xi^{(B)}\widetilde{P}_{0,k}^{D,n}+\frac{1}{2}\left(\xi^{(B)}\widetilde{P}_{0,\mathcal{M}(k)}^{A,n}+\xi^{(R)}\widetilde{P}_{\mathcal{M}(k),k}^{A,n}\right)\label{eq:PP_T}
\end{equation}
\hrulefill
\end{figure*}

In this formulation, the variables to be optimized are $s_{k}^{D,n}$,
$s_{k}^{A,n}$, $\widetilde{P}_{0,k}^{D,n}$, $\widetilde{P}_{0,\mathcal{M}(k)}^{A,n}$
and $\widetilde{P}_{\mathcal{M}(k),k}^{A,n}$, $\forall k,n$. Physically,
the constraint~(\ref{eq:C1}) ensures that the sum of the power allocated
to variables $\widetilde{P}_{0,k}^{D,n}$, $\widetilde{P}_{0,\mathcal{M}(k)}^{A,n}$
and $\widetilde{P}_{\mathcal{M}(k),k}^{A,n}$ does not exceed the
maximum power budget of the system. Constraint~(\ref{eq:C2}) ensures
that a single transmission protocol, either direct or AF, is chosen
for each user-subcarrier pair. The constraint~(\ref{eq:C3}) guarantees
that each subcarrier is only allocated to at most one user, thus intra-cell
interference is avoided. The constraints~(\ref{eq:C4}) and~(\ref{eq:C5})
describe the feasible region of the optimization variables. The OF
of problem~(P) is quasi-concave~\cite{Dinkelbach1967}, please see~\cite{Cheung2013}
for the proof.

\section{Dinkelbach's method for solving the problem (P)\label{sec:Dinkelbach}}

Dinkelbach's method~\cite{Dinkelbach1967} is an iterative algorithm
that can be used for solving a quasi-concave problem in a parameterized
concave form. The value of the parameter at iteration $i$ is denoted
by $q_{i}$, and the parameterized form is given by

Subtractive problem~(P$'$):
\begin{eqnarray}
\underset{\mathcal{P},\mathcal{S}}{\mbox{maximize}} &  & \widetilde{R}_{T}(\mathcal{P},\mathcal{S})-q_{i}\widetilde{P}_{T}(\mathcal{P},\mathcal{S})\\
\mbox{subject to} &  & (\ref{eq:C1}),(\ref{eq:C2}),(\ref{eq:C3}),(\ref{eq:C4}),(\ref{eq:C5}),\nonumber
\end{eqnarray}
which is solved at each iteration for obtaining an updated parameter
value. For futher details, please refer to~\cite{Dinkelbach1967}.

Since $\widetilde{R}_{T}(\mathcal{P},\mathcal{S})$ and $\widetilde{P}_{T}(\mathcal{P},\mathcal{S})$
are concave and affine, respectively~\cite{Cheung2013}, it is plausible
that the OF in~(P$'$) is concave. Thus,~(P$'$) is a typical concave
maximization problem and may be solved using convex optimization techniques.
In this work, we opt for the method of dual decomposition~\cite{Palomar2006},
which solves~(P$'$) by solving a series of subproblems. This approach
is favorable in this context as the OF in~(P$'$) is formed by the
summation of multiple similar terms of separate variables, where the
maximization of each term can be solved in a parallel fashion using
a decomposition technique. Let us commence by stating that the Lagrangian
of~(P$'$) is given by~(\ref{eq:lagrange}),\begin{figure*}[tbp]

\begin{eqnarray}
\mathcal{L}\left(\mathcal{P},\mathcal{S},\lambda\right) & = & \widetilde{R}_{T}(\mathcal{P},\mathcal{S})-q_{i}\widetilde{P}_{T}(\mathcal{P},\mathcal{S})+\lambda\left(P_{max}-\sum_{k=1}^{K}\sum_{n=1}^{N}\widetilde{P}_{0,k}^{D,n}+\widetilde{P}_{0,\mathcal{M}(k)}^{A,n}+\widetilde{P}_{\mathcal{M}(k),k}^{A,n}\right).\label{eq:lagrange}
\end{eqnarray}
\hrulefill
\end{figure*} where $\lambda\geq0$ is the Lagrangian multiplier associated with
the constraint~(\ref{eq:C1}). The feasible region constraints~(\ref{eq:C4})
and~(\ref{eq:C5}), as well as constraints~(\ref{eq:C2}) and~(\ref{eq:C3})
will be considered, when deriving the optimal solution, which is detailed
later.

The dual problem of~(P$'$) may be written as~\cite{Palomar2006}
$\underset{\lambda\geq0}{\mbox{min.}}\mbox{ }g(\lambda)=\underset{\lambda\geq0}{\mbox{min.}}\mbox{ }\underset{\mathcal{P},\mathcal{S}}{\mbox{max.}}\mbox{ }\mathcal{L}\left(\mathcal{P},\mathcal{S},\lambda\right)$,
which is solved by solving $NK$ similar subproblems for obtaining
both the power as well as the subcarrier allocations, and by solving
a master problem to update $\lambda$, until convergence is obtained.

\subsection{Solving the $NK$ subproblems}

These subproblems are solved by employing the Karush\textendash{}Kuhn\textendash{}Tucker~(KKT)
conditions~\cite{Boyd2004}, which are first-order necessary and
sufficient conditions for optimality. We denote all optimal variables
by a superscript asterisk, and the total transmit power assigned for
AF transmission to user $k$ over subcarrier $n$ by $\widetilde{P}_{k}^{A,n}=\widetilde{P}_{0,\mathcal{M}(k)}^{A,n}+\widetilde{P}_{\mathcal{M}(k),k}^{A,n}$.
Then, by substituting $\widetilde{P}_{\mathcal{M}(k),k}^{A,n}=\widetilde{P}_{k}^{A,n}-\widetilde{P}_{0,\mathcal{M}(k)}^{A,n}$
into~(\ref{eq:lagrange}), the following first-order derivatives
may be obtained
\begin{equation}
\left.\frac{\partial\mathcal{L}\left(\mathcal{P},\mathcal{S},\lambda\right)}{\partial\widetilde{P}_{0,k}^{D,n}}\right|_{\widetilde{P}_{0,k}^{D,n}=\widetilde{P}_{0,k}^{D,n*}}=0,\label{eq:dPb_D}
\end{equation}
\begin{equation}
\left.\frac{\partial\mathcal{L}\left(\mathcal{P},\mathcal{S},\lambda\right)}{\partial\widetilde{P}_{k}^{A,n}}\right|_{\widetilde{P}_{k}^{A,n}=\widetilde{P}_{k}^{A,n*}}=0\label{eq:dPt_A}
\end{equation}
and
\begin{equation}
\left.\frac{\partial\mathcal{L}\left(\mathcal{P},\mathcal{S},\lambda\right)}{\partial\widetilde{P}_{0,\mathcal{M}(k)}^{A,n}}\right|_{\widetilde{P}_{0,\mathcal{M}(k)}^{A,n}=\widetilde{P}_{0,\mathcal{M}(k)}^{A,n*}}=0.\label{eq:dPb_A}
\end{equation}
The optimal values of $\widetilde{P}_{0,k}^{D,n}$ may be readily
obtained from~(\ref{eq:dPb_D}) as
\begin{equation}
P_{0,k}^{D,n*}=\left[\frac{1}{\ln2\left(q_{i}\xi^{(B)}+\lambda\right)}-\frac{1}{\alpha_{k}^{D,n}}\right]^{+},\label{eq:dinkel_Pd}
\end{equation}
where the effective channel gain of the direct transmission is given
by $\alpha_{k}^{D,n}=\frac{G_{0,k}^{n}}{\Delta\gamma N_{0}W}$ and
$[\cdot]^{+}$ denotes $\max(0,\cdot)$, since the powers allocated
have to be nonnegative due to the constraint~(\ref{eq:C4}). Similarly
the optimal values of $\widetilde{P}_{0,\mathcal{M}(k)}^{A,n}$ and
$\widetilde{P}_{\mathcal{M}(k),k}^{A,n}$ may be obtained by equating~(\ref{eq:dPt_A})
and~(\ref{eq:dPb_A}) to give
\begin{equation}
P_{0,\mathcal{M}(k)}^{A,n*}=\beta_{k}^{A,n}P_{k}^{A,n*}\label{eq:dinkel_Pb}
\end{equation}
and
\begin{equation}
P_{\mathcal{M}(k),k}^{A,n*}=\left(1-\beta_{k}^{A,n}\right)P_{k}^{A,n*},\label{eq:dinkel_Pr}
\end{equation}
where the total transmit power assigned for the AF transmission to
user $k$ over subcarrier $n$ is given by~(\ref{eq:dinkel_Pt_A}),~(\ref{eq:alpha_A})
and~(\ref{eq:beta_A}).\begin{figure*}[tbp]
\begin{equation}
P_{k}^{A,n*}=\left[\frac{1}{\ln2\left(\beta_{k}^{A,n}\left(q_{i}\xi^{(B)}+2\lambda\right)+\left(1-\beta_{k}^{A,n}\right)\left(q_{i}\xi^{(R)}+2\lambda\right)\right)}-\frac{1}{\alpha_{k}^{A,n}}\right]^{+}\label{eq:dinkel_Pt_A}
\end{equation}
\begin{equation}
\alpha_{k}^{A,n}=\frac{\beta_{k}^{A,n}\left(1-\beta_{k}^{A,n}\right)G_{0,\mathcal{M}(k)}^{n}G_{\mathcal{M}(k),k}^{n}}{\left(\beta_{k}^{A,n}G_{0,\mathcal{M}(k)}^{n}+\left(1-\beta_{k}^{A,n}\right)G_{\mathcal{M}(k),k}^{n}\right)\Delta\gamma N_{0}W}\label{eq:alpha_A}
\end{equation}
\begin{equation}
\beta_{k}^{A,n}=\frac{-G_{\mathcal{M}(k),k}^{n}\left(q_{i}\xi^{(R)}+2\lambda\right)+\sqrt{G_{0,\mathcal{M}(k)}^{n}G_{\mathcal{M}(k),k}^{n}\left(q_{i}\xi^{(B)}+2\lambda\right)\left(q_{i}\xi^{(R)}+2\lambda\right)}}{G_{0,\mathcal{M}(k)}^{n}\left(q_{i}\xi^{(B)}+2\lambda\right)-G_{\mathcal{M}(k),k}^{n}\left(q_{i}\xi^{(R)}+2\lambda\right)}\label{eq:beta_A}
\end{equation}
\hrulefill
\end{figure*} Observe that~(\ref{eq:beta_A}) is the fraction of the total AF
transmit power that is allocated for the BS-to-RN link while obeying
$0\leq\beta_{k}^{A,n}\leq1$.

Having calculated the optimal power allocations, the optimal subcarrier
allocations may be derived using the first-order derivatives as follows:
\begin{eqnarray}
\frac{\partial\mathcal{L}\left(\mathcal{P},\mathcal{S},\lambda\right)}{\partial s_{k}^{D,n}} & = & \log_{2}\left(1+\alpha_{k}^{D,n}P_{0,k}^{D,n*}\right)\nonumber \\
 &  & -\frac{\alpha_{k}^{D,n}P_{0,k}^{D,n*}}{\ln2\left(1+\alpha_{k}^{D,n}P_{0,k}^{D,n*}\right)}\nonumber \\
 & = & D_{k}^{n}\begin{cases}
<0 & \mbox{if }s_{k}^{D,n*}=0,\\
=0 & \mbox{if }s_{k}^{D,n*}\in(0,1)\\
>0 & \mbox{if }s_{k}^{D,n*}=1
\end{cases},\label{eq:dinkel_Dnk}
\end{eqnarray}
and
\begin{eqnarray}
\frac{\partial\mathcal{L}\left(\mathcal{P},\mathcal{S},\lambda\right)}{\partial s_{k}^{A,n}} & = & \frac{1}{2}\log_{2}\left(1+\alpha_{k}^{A,n}\widetilde{P}_{k}^{A,n*}\right)\nonumber \\
 &  & -\frac{\alpha_{k}^{A,n}\widetilde{P}_{k}^{A,n*}}{2\ln2\left(1+\alpha_{k}^{A,n}\widetilde{P}_{k}^{A,n*}\right)}\\
 & = & A_{k}^{n}\begin{cases}
<0 & \mbox{if }s_{k}^{A,n*}=0,\\
=0 & \mbox{if }s_{k}^{A,n*}\in(0,1)\\
>0 & \mbox{if }s_{k}^{A,n*}=1.
\end{cases},\label{eq:dinkel_Ank}
\end{eqnarray}
(\ref{eq:dinkel_Dnk}) and~(\ref{eq:dinkel_Ank}) stem from the fact
that if the optimal value of $s_{k}^{X,n}$ occurs at the boundary
of the feasible region, then $\mathcal{L}\left(\mathcal{P},\mathcal{S},\lambda\right)$
must be decreasing with the values of $s_{k}^{X,n}$ that approach
the interior of the feasible region. By contrast, for example, the
derivative $D_{k}^{n}=0$ if the optimal $s_{k}^{D,n}$ is obtained
in the interior of the feasible region~\cite{Wong1999}. However,
since each subcarrier may only be used for transmission to a single
user, each subcarrier $n$ is allocated to the specific user $k$
having the highest value of $\max\left(A_{k}^{n},D_{k}^{n}\right)$
in order to achieve the highest increase in $\mathcal{L}\left(\mathcal{P},\mathcal{S},\lambda\right)$.
The optimal allocation for subcarrier $n$ is as follows%
\footnote{If there are multiple users that tie for the maximum $\max\left(A_{k}^{n},D_{k}^{n}\right)$,
a random user from the maximal set is chosen. %
}
\begin{equation}
s_{k}^{D,n*}=\left\{ \begin{array}{cc}
1, & \mbox{if }D_{k}^{n}=\max_{j}\left[\max\left(A_{j}^{n},D_{j}^{n}\right)\right]\mbox{ and }D_{k}^{n}\geq0,\\
0, & \mbox{otherwise,}
\end{array}\right.\label{eq:dinkel_sd}
\end{equation}
and

\begin{equation}
s_{k}^{A,n*}=\left\{ \begin{array}{cc}
1, & \mbox{if }A_{k}^{n}=\max_{j}\left[\max\left(A_{j}^{n},D_{j}^{n}\right)\right]\mbox{ and }A_{k}^{n}\geq0,\\
0, & \mbox{otherwise.}
\end{array}\right.\label{eq:dinkel_sa}
\end{equation}
Thus constraints~(\ref{eq:C2})-~(\ref{eq:C5}) are satisfied and
the optimal primal variables are obtained for a given $\lambda$.
Observe that the optimal power allocations given by~(\ref{eq:dinkel_Pd})
and~(\ref{eq:dinkel_Pt_A}) are indeed customized water-filling solutions,
where the effective channel gains are given by $\alpha_{k}^{D,n}$
and $\alpha_{k}^{A,n}$, respectively, and where the water levels
are determined both by the cost of allocating power, $\lambda$, as
well as the current cost of power to the EE given by $q_{i}$.

\subsection{Updating the dual variable $\lambda$}

\begin{figure}[h]
\begin{centering}
\includegraphics[scale=0.7]{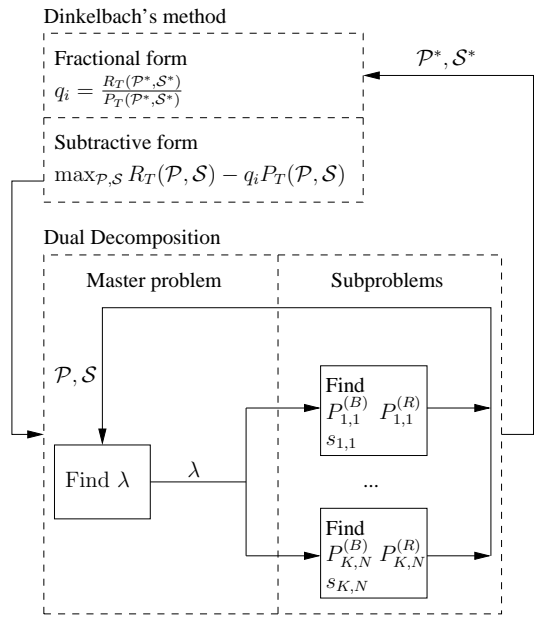}
\par\end{centering}

\caption{Summary of the solution methodology for the relaxed problem (P).}
\label{fig:summary}
\end{figure}

Since~(\ref{eq:dinkel_Pd}),~(\ref{eq:dinkel_Pb}),~(\ref{eq:dinkel_Pr}),~(\ref{eq:dinkel_sd})
and~(\ref{eq:dinkel_sa}) give a unique solution for $\mbox{\ensuremath{\underset{\mathcal{P},\mathcal{S}}{\mbox{max.}}} }\mathcal{L}\left(\mathcal{P},\mathcal{S},\lambda\right)$,
it follows that $g(\lambda)$ is differentiable and hence the gradient
method~\cite{Boyd2004,Palomar2006} may be readily used for updating
the dual variables $\lambda$. The gradient of $\lambda$ is given
by
\begin{equation}
\frac{\partial\mathcal{L}\left(\mathcal{P},\mathcal{S},\lambda\right)}{\partial\lambda}=P_{max}-\sum_{k=1}^{K}\sum_{n=1}^{N}\Big(\widetilde{P}_{0,k}^{D,n}+\widetilde{P}_{0,\mathcal{M}(k)}^{A,n}+\widetilde{P}_{\mathcal{M}(k),k}^{A,n}\Big).
\end{equation}
Therefore, $\lambda$ may be updated using the optimal variables to
give~(\ref{eq:Dinkel_lambda}),\begin{figure*}[tbp]
\begin{equation}
\lambda(i+1)=\left[\lambda(i)-\alpha_{\lambda}(i)\left(P_{max}-\sum_{k=1}^{K}\sum_{n=1}^{N}\widetilde{P}_{0,k}^{D,n*}+\widetilde{P}_{0,\mathcal{M}(k)}^{A,n*}+\widetilde{P}_{\mathcal{M}(k),k}^{A,n*}\right)\right]^{+}\label{eq:Dinkel_lambda}
\end{equation}
\hrulefill
\end{figure*} where $\alpha_{\lambda}(i)$ is the size of the step taken in the
direction of the negative gradient for the dual variable $\lambda$
at iteration $i$. For the performance investigations of Section~\ref{sec:Results-and-Discussions},
a constant step size is used, since it is comparatively easier to
find a value that strikes a balance between optimality and convergence
speed. The process of computing the optimal power as well as subcarrier
allocations and subsequently updating $\lambda$ is repeated until
convergence is attained, indicating that the dual optimal point has
been reached. Since the primal problem~(P$'$) is concave, there
is zero duality gap between the dual and primal solutions. Hence,
solving the dual problem is equivalent to solving the primal problem.
For additional clarity, the solution methodology is summarized in
Fig.~\ref{fig:summary}.

\section{Numerical results and discussions\label{sec:Results-and-Discussions}}

\begin{table}
\begin{centering}
\caption{Simulation parameters used to obtain all results in this section unless
otherwise specified.}
\label{tab:param}%
\begin{tabular}{|l|r|}
\hline
Simulation parameter & Value\tabularnewline
\hline
\hline
Subcarrier bandwidth, $W$ Hertz & $12$k\tabularnewline
\hline
$P_{C}^{(B)}$ Watts~\cite{Arnold2010} & 60\tabularnewline
\hline
$P_{C}^{(R)}$ Watts~\cite{Arnold2010} & 20\tabularnewline
\hline
$\xi^{(B)}$~\cite{Arnold2010} & 2.6\tabularnewline
\hline
$\xi^{(R)}$~\cite{Arnold2010} & 5\tabularnewline
\hline
$N_{0}$ dBm/Hz & \textminus{}174\tabularnewline
\hline
Convergence tolerance & $10^{-8}$\tabularnewline
\hline
Number of channel samples & $10^{4}$\tabularnewline
\hline
\end{tabular}
\par\end{centering}

\vspace{-5mm}
\end{table}
This section presents the results of applying the EEM algorithm described
in Section~\ref{sec:Dinkelbach} to the relay-aided cellular system
shown in Fig.~\ref{fig:cellular}, where the RNs are placed halfway
between the BS and the cell-edge. The channel is modeled by the path-loss~\cite{3GPP_PL}
and uncorrelated Rayleigh fading obeying the complex normal distribution,
$\mathcal{CN}(0,1)$. It is assumed that the BS-to-RN link has line-of-sight~(LOS)
propagation, implying that a RN was placed on a tall building. However,
the BS-to-UE and RN-to-UE links typically have no LOS, since these
links are likely to be blocked by buildings and other large obstructing
objects. An independently-random set of UE locations as well as fading
channel realizations are generated for each channel sample. For fair
comparisons, the metrics used are the average SE per subcarrier and
the average EE per subcarrier. On the other hand, the sum-rate may
be calculated by multiplying the average SE by $NW.$ Additionally,
$\rho$ is introduced to denote the average fraction of the total
number of subcarriers that are used for AF transmission. Thus, $\rho$
quantifies the benefit attained from introducing RNs into the system.

\begin{figure}
\begin{centering}
\includegraphics[scale=0.8]{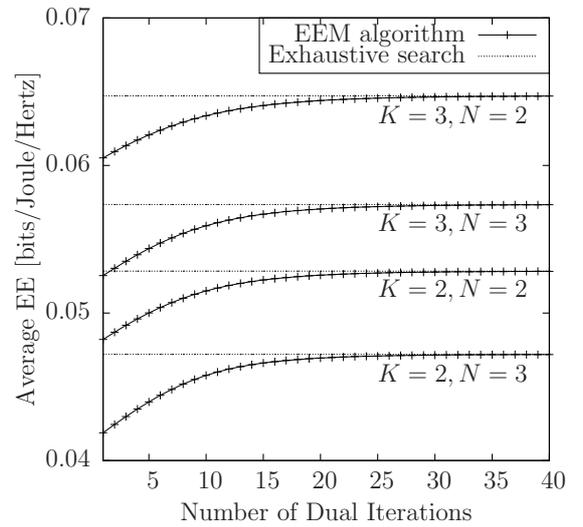}
\par\end{centering}

\caption{Average EE versus the total number of inner iterations of Dinkelbach's
method required for reaching convergence when using the simulation
parameters from Table~\ref{tab:param} with $P_{max}=0$dBm, $M=0$
and with a cell radius of $1$km.}
\label{fig:conv_main}
\end{figure}
Fig.~\ref{fig:conv_main} illustrates the convergence behavior of
Dinkelbach's method invoked for maximizing the EE for a selection
of small-scale systems, averaged over $10^{4}$ different channel
realizations. Since the problem size is small, it is possible to generate
also the exhaustive-search based solution within a reasonable computation
time. As seen in Fig.~\ref{fig:conv_main}, Dinkelbach's method converges
to the optimal value within forty inner iterations. This result demonstrates
that the EEM algorithm based on Dinkelbach's method indeed obtains
the optimal power and subcarrier allocation, even though the relaxed
problem is solved and a high receiver's SNR was assumed.

\begin{figure}
\begin{centering}
\subfloat[Average SE and $\rho$ versus $P_{max}$ for $K=30$, $60$ and $120$.]{\begin{centering}
\includegraphics[scale=0.8]{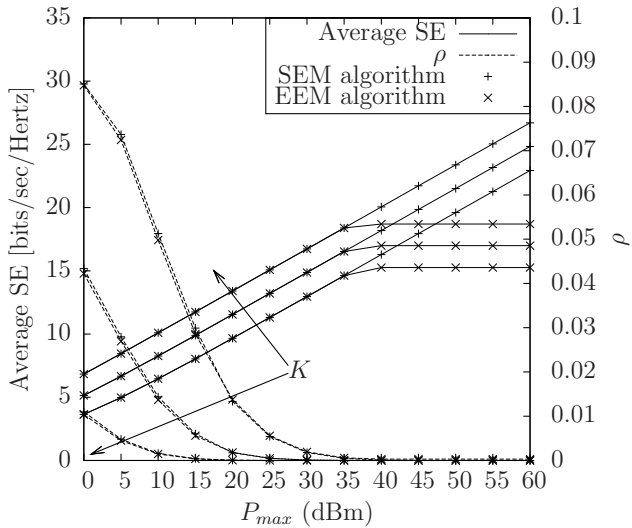}
\par\end{centering}

\label{fig:users_SE}}
\par\end{centering}

\begin{centering}
\subfloat[Average EE and $\rho$ versus $P_{max}$ for $K=30$, $60$ and $120$.]{\begin{centering}
\includegraphics[scale=0.8]{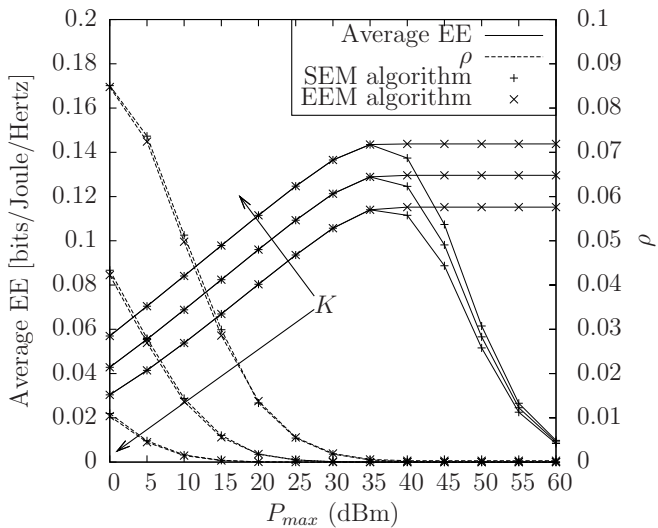}
\par\end{centering}

\label{fig:users_EE}}
\par\end{centering}

\centering{}\caption{Average SE, EE and $\rho$, and the effect of an increasing number
of users, $K$, for a system with simulation parameters from Table~\ref{tab:param}
with $N=128$, $M=3$ and with a cell radius of $1.5$km.}
\label{fig:users}
\end{figure}
Additionally, the EEM algorithm may be employed for evaluating the
effects of system-level design choices on the network's SE and EE.
The effect of $K$ on the average EE and SE%
\footnote{N.B. The maximum SE is obtained in the first outer iteration of Dinkelbach's
method with $q_{0}=0$, since this equates to zero penalty for any
power consumption.%
} is depicted in Fig.~\ref{fig:users}. As expected, upon increasing
$K$, the multi-user diversity of the system is increased, since the
scheduler is allowed to choose its subcarrier allocations from a larger
pool of channel gains. This results in an increase of both the maximum
EE as well as of the SE attained. Furthermore, Fig.~\ref{fig:users}
shows that as $P_{max}$ is increased, the SEM algorithm continues
to allocate more power in order to achieve a higher average SE at
the cost of EE, while the EEM algorithm attains the maximum EE and
does not continue to increase its attainable SE by sacrificing the
achieved EE. On the other hand, $\rho$ is inversely proportional
to $K$. This indicates that as the multi-user diversity increases,
the subcarriers are less likely to be allocated for AF transmissions,
simply because there are more favorable BS-to-UE channels owing to
having more UEs nearer to the cell-center. Moreover, the value of
$\rho$ decreases as $P_{max}$ increases, because there is more power
to allocate to the BS-to-UE links for UEs near the cell-center, which
benefit from a reduced pathloss as well as from a more efficient power
amplifier at the BS.

\section{Conclusions\label{sec:Conclusions}}

In this paper, the joint power and subcarrier allocation problem was
formulated for maximizing the EE in a multi-relay aided multi-user
OFDMA cellular network. Through an introduction of auxiliary variables
and a relaxation of the binary-constrained variables, the OF of the
problem can be shown to be quasi-concave. Thus, Dinkelbach's method
is employed for solving the problem and we have shown that this solution
method obtains the same solutions as an exhaustive search and is therefore
optimal. Additionally, we analyze the effect of perturbing the number
of available UEs on the system's achievable SE and EE.\bibliographystyle{IEEEtran}
\bibliography{references}

\end{document}